\documentclass{aa}
\usepackage{graphicx}
\usepackage{multirow}
\usepackage{multicol}

\def\ms{\ifmmode {\rm M_{\odot}} \else ${\rm M_{\odot}}$\fi}    

\begin{document}

\title{The kHz QPOs as a probe of the   X-ray color-color diagram and  accretion-disk structure for the atoll source 4U 1728-34}

\author{De-Hua Wang\inst{1}\thanks{wangdh@gznu.edu.cn(DHW); zhangcm@bao.ac.cn(CMZ)}
\and Cheng-Min Zhang\inst{2,3,4\star}
\and Jin-Lu Qu\inst{5}
}


\institute{School of Physics and Electronic Science, Guizhou Normal University, Guiyang, 550001, China\\\email{wangdh@gznu.edu.cn}
\and National Astronomical Observatories, Chinese Academy of Sciences, Beijing, 100101, China\\\email{zhangcm@bao.ac.cn}
\and School of Physical Sciences, University of Chinese Academy of Sciences, Beijing 101400, China
\and CAS Key Laboratory of FAST, Chinese Academy of Sciences, Beijing 100101, China
\and Institute of High Energy Physics, Chinese Academy of Sciences, Beijing, 100049, China
}

\date{Received 2 November 1992 / Accepted 7 January 1993}

\abstract{
We have taken the kHz QPOs as a tool to probe the correlation between the tracks of X-ray color-color diagram (CCD) and   magnetosphere-disk  positions for the atoll source 4U 1728-34, based on the assumptions that the upper kHz QPO is ascribed to the Keplerian orbital motion and  the neutron star (NS) magnetosphere is defined by the dipole magnetic field. We find that from the island to the banana state, the inner accretion disk gradually  approaches the NS surface with the radius decreasing from $r\sim33.0$\,km to $\sim15.9$\,km, corresponding to the magnetic field from $B(r)\sim4.8\times10^6$\,G to $\sim4.3\times10^7$\,G.
In addition,   we note the characteristics of some particular radii of  magnetosphere-disk --$r$ are:
firstly, the whole atoll shape of the CCD links the disk radius range of $\sim15.9-33.0$\,km, which is just located inside
  the corotation radius of 4U 1728-34 ---$r_{\rm co}$ ($\sim34.4$\,km), implying that the CCD shape   is involved in the
 NS  spin-up state.
Secondly, the   island and banana states of CCD  correspond to the two particular boundaries: (I)---near the corotation  radius at $r\sim27.2-33.0$\,km,
where the source lies  in the island state; (II)---near the NS surface at $r\sim15.9-22.3$\,km,
where the source lies  in both the island and banana states.
Thirdly, the  vertex of the atoll shape in CCD, where the radiation transition  from the hard to soft photons occurs,
is found to be near the NS surface at $r\sim16.4$\,km.
The above results suggest that both the  magnetic field and accretion environment are related to the CCD structure of atoll track, where the corotation radius and NS hard surface play the significant roles in the radiation distribution of atoll source.
}


\authorrunning{De-Hua Wang et al.}

\titlerunning{The kHz QPOs as a probe of the ...}

\maketitle

\section{Introduction} \label{sec:intro}

Neutron star low mass X-ray binaries (NS-LMXBs) can be classified as atoll and Z sources \citep{Hasinger89,van der Klis06} according to the patterns they trace out in the X-ray color-color diagrams (CCDs). From the top to bottom in CCD, the tracks of Z sources are called the horizontal, normal, and flare branches, while for atoll sources, they are called the extreme island, island, lower banana, and upper banana states. Initially, it is assumed that the mass-accretion rate of Z sources may increase monotonically along the branches \citep{Hasinger90b,Hasinger90a}. However, this scenario cannot be used to interpret the "parallel tracks" phenomenon, in other words, the similar temporal and spectral properties are observed at the different luminosity \citep{Mendez99,Ford00,van der Klis01}.
Alternatively, \citet{van der Klis01} proposed that the changes in the averaged accretion rate
through the disk could be responsible for the motion along the Z track.
In addition, the transition between the atoll and Z tracks has been observed in XTE J1701-462, which is explained as the influence by the changes in the accretion rate \citep{Homan07,Lin09,Homan10,Fridriksson15}.
Other physical effects, for example, the NS magnetic field \citep{Hasinger89}, have been proposed to interpret the difference between the
two-subclass sources, however, there is no consensus as to their origin.

\begin{figure}
\centering
\includegraphics[width=9cm]{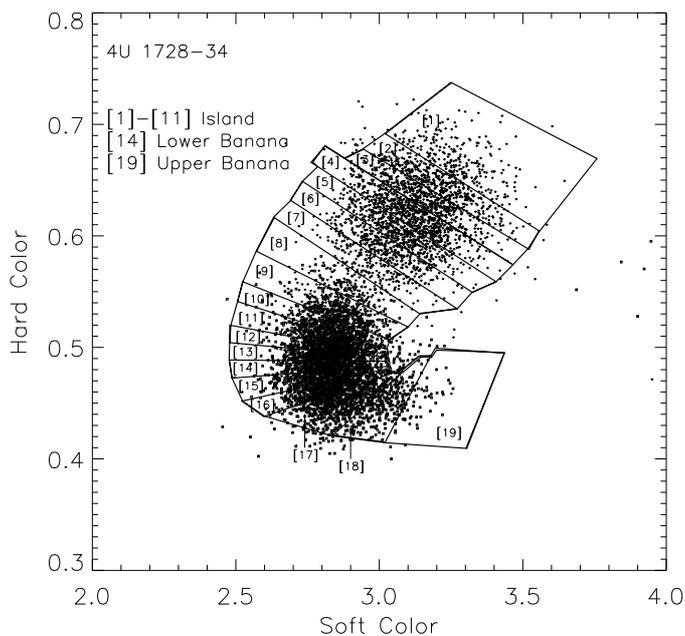}
\caption{X-ray color-color diagram of 4U 1728-34,
where the soft and hard colors are defined as the ratio of the count rates in the bands $3.5-6.4$\,keV/$2.0-3.5$\,keV
and $9.7-16$\,keV/$6.4-9.7$\,keV, respectively. This labeled area of boxes indicate the intervals used to select the power spectra. The figure is reproduced with the same dataset of $RXTE$ used by \citet{Di Salvo01}.
}
\label{origion}
\end{figure}

In both the atoll and Z sources the twin kilohertz quasi-periodic oscillations (kHz QPOs, with lower $\nu_1$ and upper $\nu_2$) have been detected by $Rossi$ $X-ray$ $Timing$ $Explorer$ ($RXTE$) \citep{van der Klis06}, which show a non-linear relation between $\nu_1$ and $\nu_2$ \citep{Belloni02,Belloni05,Zhang06a,Belloni07} in the frequency range of $\simeq100-1300$\,Hz \citep{Wang14}.
The frequencies of kHz QPOs are found to be correlated  with other temporal and spectral features (e.g., \citealt{Ford98,Kaaret98,Psaltis99,Belloni02,Mendez06}), and they often increase monotonically along both the atoll  tracks
(e.g., \citealt{van Straaten00,van Straaten03,Altamirano08b}) and Z tracks (e.g., \citealt{Wijnands97,Wijnands98,Homan02,Jonker02,Lin12}).
It is suggested that kHz QPOs may reflect the orbital  motion of the accreting matter  at the inner accretion disk (\citealt{Kluzniak90};
\citealt{van der Klis00} and references therein; \citealt{Zhang13}),
which can be exploited to explore the physical environments in the strong gravitational field and strong
magnetic field regions \citep{Kluzniak90,Abramowicz03b,Miller15}.
\citet{Wang15,Wang17} analyzed the relations of the kHz QPO emission radii with the NS radius and corotation radius, suggesting
that the emission of kHz QPOs may be affected by the NS hard surface and the spin-up environment near the corotation radius.

As yet, we do not understand why some atoll sources have been detected with the millisecond spin signals but the Z sources have not. In addition, among $\sim30$ NS-LMXBs with detected NS spin signals \citep{Burderi13,van der Klis16},
$12$ sources have also been detected with twin kHz QPOs \citep{Wang17,Wang18}, including ten atoll sources.
Particularly, in the atoll source 4U 1728-34 the  X-ray burst oscillation has been observed with the
frequency of 363 Hz \citep{Strohmayer96,Zhang16}, while it has also shown temporal variabilities on all timescales including
kHz QPOs (e.g., \citealt{Mendez99a,Di Salvo01,van Straaten02,Mukherjee12}) and spectral evolution (e.g., \citealt{Tarana11,Mondal17}).
The goal of this paper is to take the kHz QPOs  as the tool to probe the relation between the X-ray CCD tracks of 4U 1728-34  and the magnetosphere-disk positions, and further investigate the effects of accretion environment and NS magnetic field on the formation of the atoll track.

The paper is organized as follows: In $\S$ 2, we introduce the data of twin kHz QPOs and NS spin of 4U 1728-34. In $\S$ 3
we probe the magnetosphere-disk positions and magnetic field strength along the atoll track. We have investigated the relation between
the CCD shape and the particular disk radii, for example, NS radius and its corotation radius.
In $\S$ 4 we present the discussions and conclusions.

\begin{table*}
\begin{minipage}{180mm}
\centering
\caption{kHz QPOs and neutron star spin of 4U 1728-34 and the inferred parameters.}
\begin{tabular}{@{}lcccccccc@{}}
\hline
\hline
\noalign{\smallskip}
Interval & $\nu_1^a$ & $\nu_2^b$ & $\nu_{\rm s}^c$ & $r^d$ & $B(r)^e$ & $r_{\rm co}^f$ \\
Number & (Hz) & (Hz) & (Hz) & (km)  & (G) & (km) \\
\noalign{\smallskip}
\hline
\noalign{\smallskip}
[1]~\ldots\ldots & \ldots & $387\pm18$ & 363 & $33.0$ & $4.8\times10^6$ & 34.4 \\\relax
[2]~\ldots\ldots & \ldots & $397\pm14$ & & $32.4$ & $5.1\times10^6$ & \\\relax
[3]~\ldots\ldots & \ldots & $466\pm21$ & & $29.1$ & $7.0\times10^6$ & \\\relax
[4]~\ldots\ldots & \ldots & $497.8\pm5.3$ & & $27.9$ & $8.0\times10^6$ & \\\relax
[5]~\ldots\ldots & \ldots & $498.7\pm5.9$ & & $27.9$ & $8.0\times10^6$ & \\\relax
[6]~\ldots\ldots & \ldots & $517.3\pm8.1$ & & $27.2$ & $8.6\times10^6$ & \\\relax
[7]~\ldots\ldots & \ldots & $694.5\pm9.3$ & & $22.3$ & $1.5\times10^7$ & \\\relax
[8]~\ldots\ldots & \ldots & $729.0\pm3.5$ & & $21.6$ & $1.7\times10^7$ & \\\relax
[9]~\ldots\ldots & \ldots & $789.6\pm2.5$ & & $20.5$ & $2.0\times10^7$ & \\\relax
[10]~\ldots\ldots & $511\pm17$ & $846.9\pm2.0$ & & $19.6$ & $2.3\times10^7$ & \\\relax
[11]~\ldots\ldots & $559\pm11$ & $872.9\pm1.6$ & & $19.2$ & $2.4\times10^7$ & \\\relax
[12]~\ldots\ldots & $599\pm14$ & $905.2\pm2.6$ & & $18.7$ & $2.6\times10^7$ & \\\relax
[13]~\ldots\ldots & $673\pm10$ & $948.8\pm3.8$ & & $18.1$ & $2.9\times10^7$ & \\\relax
[14]~\ldots\ldots & $750.9\pm3.7$ & $1054.7\pm8.7$ & & $16.9$ & $3.6\times10^7$ & \\\relax
[15]~\ldots\ldots & $773.0\pm1.8$ & $1105.8\pm6.8$ & & $16.4$ & $3.9\times10^7$ & \\\relax
[16]~\ldots\ldots & $816.0\pm4.0$ & $1129\pm13$ & & $16.2$ & $4.1\times10^7$ & \\\relax
[17]~\ldots\ldots & $876.1\pm2.6$ & $1158\pm18$ & & $15.9$ & $4.3\times10^7$ & \\
\noalign{\smallskip}
\hline
\noalign{\smallskip}
\end{tabular}
\label{QPOs}
\end{minipage}
\begin{minipage}{180mm}
$^{a}$ $\nu_1$---Frequency of the lower kHz QPO from \citet{Di Salvo01}.
$^{b}$ $\nu_2$---Frequency of the upper kHz QPO from \citet{Di Salvo01}.
$^{c}$ $\nu_{\rm s}$---Frequency of the neutron star spin from \citet{Strohmayer96}.
$^{d}$ $r$---Emission radius of the kHz QPOs (the magnetosphere-disk radius) inferred by equation (\ref{r_K}) with the assumed NS mass  $M\sim1.6\,\rm M_\odot$.
$^{e}$ $B(r)$---Magnetic field strength at the magnetosphere-disk radius inferred by equation (\ref{B}) with the assumed NS surface magnetic field strength $B_{\rm s}\sim10^8$\,G and NS radius $R\sim12$\,km.
$^{f}$ $r_{\rm co}$---Corotation radius inferred by equation (\ref{r_co}) with the assumed NS mass $M\sim1.6\,\rm M_\odot$.
\end{minipage}
\label{QPO}
\end{table*}

\section{Parameters of 4U 1728-34}

We collected the kHz QPO frequencies of 4U 1728-34 detected by \citet{Di Salvo01} and the NS spin frequency detected by \citet{Strohmayer96}.
For the kHz QPOs, \citet{Di Salvo01} analyzed the $RXTE$ data of 4U 1728-34 in 1996 between February 15 and March 1, 1996 on May 3,
and 1997 between September 23 and October 1, with a total observation time of  $\sim456$\,ks. The authors plot the CCD by defining the soft and hard colors in CCD  as the ratio of the count rate in the bands $3.5-6.4$\,keV/$2.0-3.5$\,keV and $9.7-16$\,keV/$6.4-9.7$\,keV, respectively (see Figure \ref{origion} in which we reproduce the CCD with the same dataset of $RXTE$ used by \citet{Di Salvo01}).
Then they divided the CCD into 19 intervals by the boxes as shown in Figure \ref{origion} (see also \citealt{Di Salvo01}), and further computed a power spectrum for each interval and fit the QPOs.
The kHz QPOs have been detected in the intervals from [1] to [17] with the range of $\nu_1\sim511-876.1$\,Hz and $\nu_2\sim387-1158$\,Hz (see Table \ref{QPOs}), where it can be seen that $\nu_2$ increases  monotonically along the atoll track
from the interval [1] to [17]. According to \citet{Di Salvo01}, the power spectra of interval [1]-[11] are typical of the island state of atoll sources, while the intervals [14]-[19] correspond to the lower and  upper banana state, respectively.
For the NS spin, \citet{Strohmayer96} detected the burst oscillation of 4U 1728-34 with the frequency of 363\,Hz during the type I X-ray burst (see Table \ref{QPOs}), which can be inferred as the NS spin frequency \citep{Boutloukos08a}.

\section{Characteristic radii and magnetic field strength}

It is generally  thought that the kHz QPOs reflect the motion of the accreting matter at the inner disk boundary around NS (\citealt{Kluzniak90};
\citealt{van der Klis00,van der Klis06} and references therein).
This implies the emission radius of the kHz QPOs may indicate the magnetosphere-disk radius.
Furthermore, the kHz QPOs and color parameters at the different intervals in Figure \ref{origion} are calculated
by the same light curves (see \citealt{Di Salvo01} for the details),
so the magnetosphere-disk structures at various positions in X-ray CCD can be approximately
inferred by the emission radius of kHz QPOs.
In this paper, we take the detected kHz QPOs \citep{Di Salvo01} and NS spin \citep{Strohmayer96} of 4U 1728-34 to investigate the evolution of its magnetosphere-disk structure along the atoll track based on some assumptions (see the following equations), including the Keplerian orbital motion and magnetic dipole field.

\subsection{Magnetosphere-disk radius inferred by kHz QPOs}

Usually, the upper kHz QPO frequency $\nu_2$ is assumed as the Keplerian orbital frequency $\nu_{\rm K}$ of the accretion plasma. The emission radius of the kHz QPOs is explained as the magnetosphere-disk radius (e.g., \citealt{Miller98,Stella99a,Stella99b,Lamb01,Zhang04}):
\begin{equation}
\nu_2=\nu_{\rm K}=\sqrt{\frac{GM}{4{\rm \pi}^2r^3}},
\label{nu_K}
\end{equation}
where $G$ is the gravitational constant, $M$ is the NS mass and $r$ is the emission radius of the kHz QPOs
referring to the NS center, that is, the magnetosphere-disk radius. By solving equation (\ref{nu_K}), $r$ can be derived as
\begin{equation}
\begin{aligned}
r&=(\frac{GM}{4{\rm \pi}^2})^{1/3}\nu_2^{-2/3}\\
&\approx18.8({\rm km})(\frac{M}{1.6\,{\rm M_\odot}})^{1/3}(\frac{\nu_2}{900\,{\rm Hz}})^{-2/3}.
\end{aligned}
\label{r_K}
\end{equation}

We infer the magnetosphere-disk radius---$r$ by equation (\ref{r_K}) with the detected $\nu_2$ values in Table \ref{QPOs} and the assumed NS mass
of $M\sim1.6\,{\rm M_\odot}$ (on the average mass of the millisecond pulsars, see \citealt{Zhang11} and \citealt{Ozel16}).
The results are shown in Table \ref{QPOs}, where it is noticed  that $r$ lies  in the range of $\sim15.9-33.0$\,km and it decreases monotonically from
the internal [1] ($r\sim33.0$\,km) to interval [17] ($r\sim15.9$\,km), which decreases by a factor of $\sim50$\%.

Figure \ref{r_b_mdot_rho} (a) shows the plot of the magnetosphere-disk radius corresponding to each interval in CCD, where $r$ decreases monotonically along the atoll track, from the island state to the banana state. It should be also noticed that there is a radius gap between the interval [6] ($r\sim27.2$\,km) and interval [7] ($r\sim22.3$\,km)
with few photons, and there is a vertex around the internal [15] ($r\sim16.4$\,km) with the radiation changing from the
hard photons dominated gradually to the soft photons dominated. For clarity, Figure \ref{schematic_diagram_r_r_co} shows the schematic diagram
of the magnetosphere-disk structures corresponding to the internal [1]-[17] in Figure \ref{r_b_mdot_rho} (a), where the $\sim5$\,km gap of the magnetosphere-disk radius between the internal [6] ($r\sim27.2$\,km) and [7] ($r\sim22.3$\,km) is obviously noticed.

\subsection{Magnetic field strength at the magnetosphere-disk radius}
\begin{figure}
\centering
\includegraphics[width=9cm]{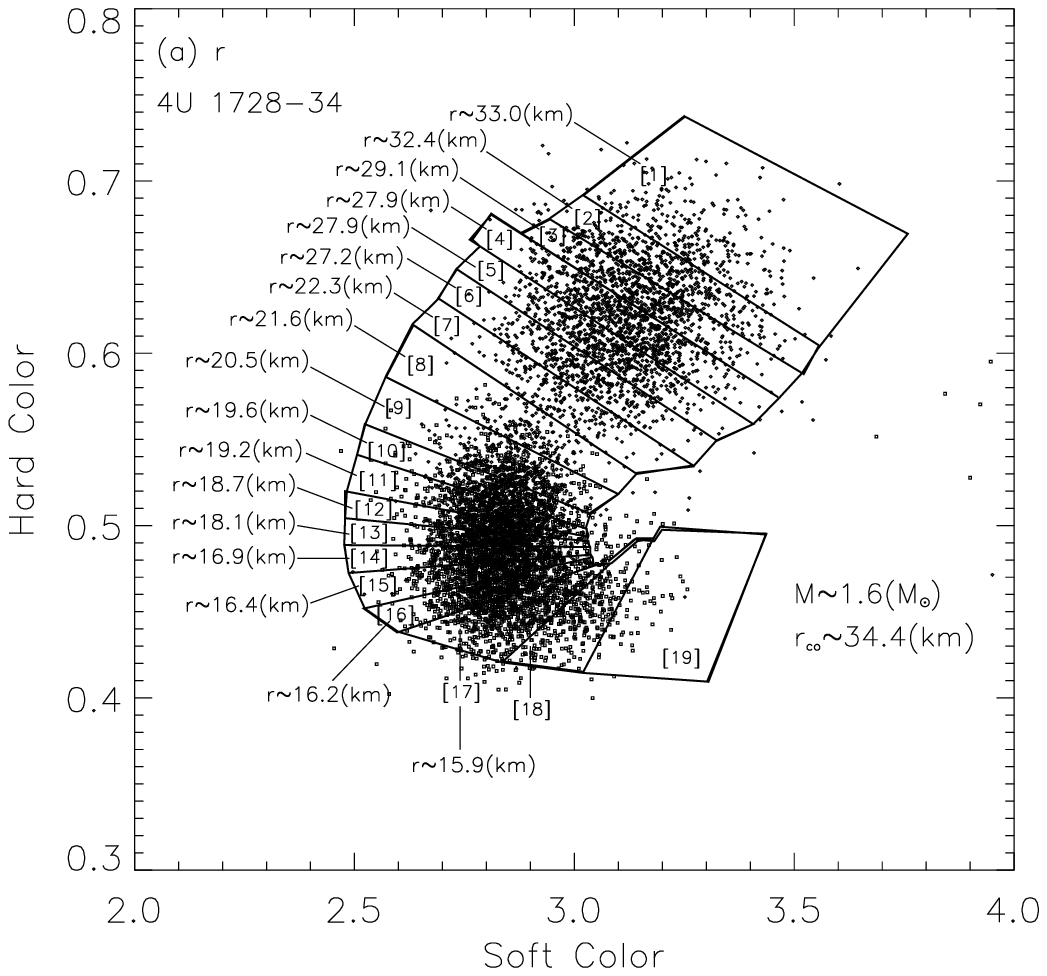}
\includegraphics[width=9cm]{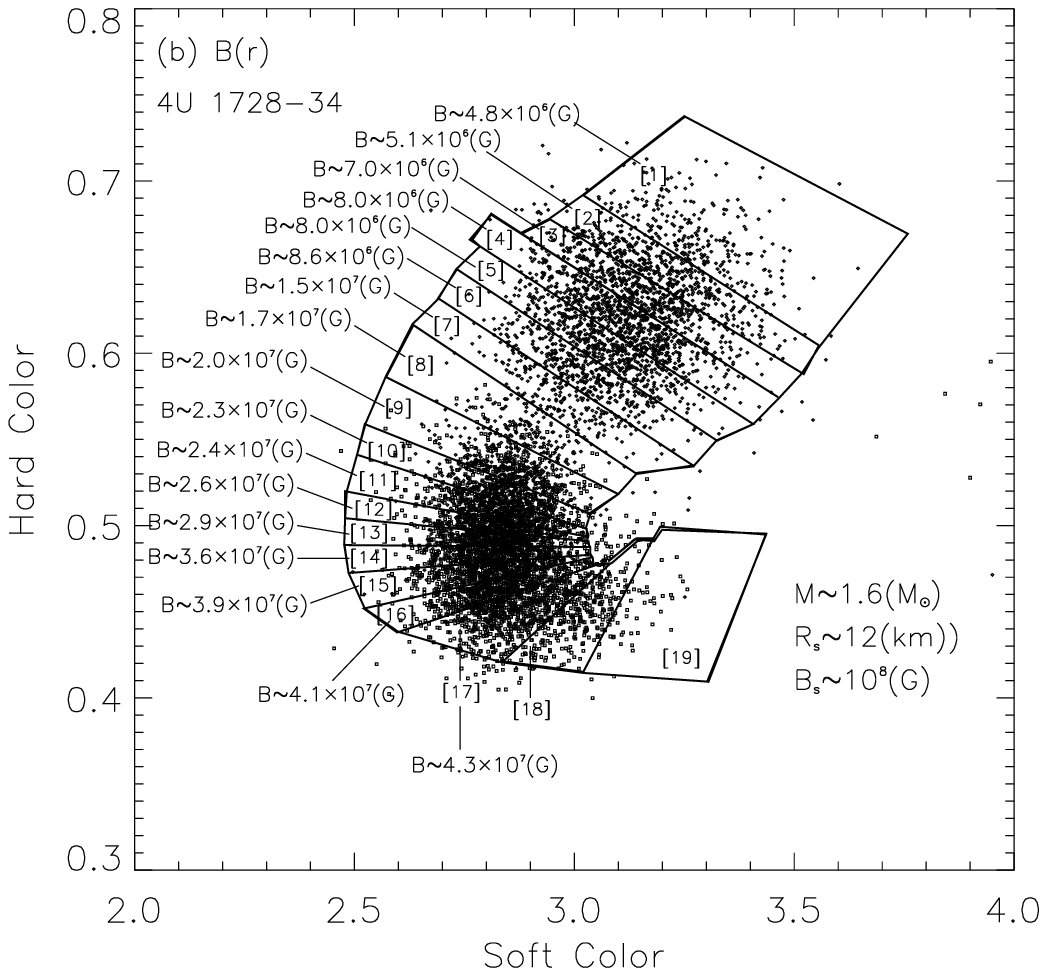}
\caption{(a)---Magnetosphere-disk radius $r$ at various positions (interval [1]-[17]) in X-ray color-color diagram. (b)---Similar to (a),
but for the magnetic field strength $B$ at $r$.
The observed CCD points of 4U 1728 -34 are taken from the paper by \citet{Di Salvo01}. The figure is reproduced with the $RXTE$ dataset used by \citet{Di Salvo01}.
}
\label{r_b_mdot_rho}
\end{figure}

In this paper the magnetic field of NS is assumed to be dipolar \citep{Bhattacharya91}:
\begin{equation}
\begin{aligned}
B(r)&=B_{\rm s}(\frac{R}{r})^3\\
&\propto r^{-3},
\end{aligned}
\label{B}
\end{equation}
where $r$ is the radial distance referring to the NS center, $B_{\rm s}$ and $R$ are the surface magnetic field strength and the stellar radius of NS, respectively.

We inferred the magnetic field strength at the magnetosphere-disk radius, i.e., $B(r)$, by equation (\ref{B}) with the inferred emission radius
of kHz QPOs in Table \ref{QPOs}, as well as with the assumed NS surface magnetic field strength $B_{\rm s}\sim10^8$\,G \citep{Zhang06,Harding13,Mondal17} and NS radius $R\sim12$\,km \citep{Shaposhnikov03,Miller15}.
The results are shown in Table \ref{QPOs}, where $B(r)$  ranges at $\sim4.8\times10^6\,{\rm G}-4.3\times10^7\,{\rm G}$, and it increases
monotonically from the internal [1] ($B(r)\sim4.8\times10^6$\,G) to interval [17] ($B(r)\sim4.3\times10^7$\,G),  by nearly one order of magnitude.

\begin{figure}
\centering
\includegraphics[width=8.2cm]{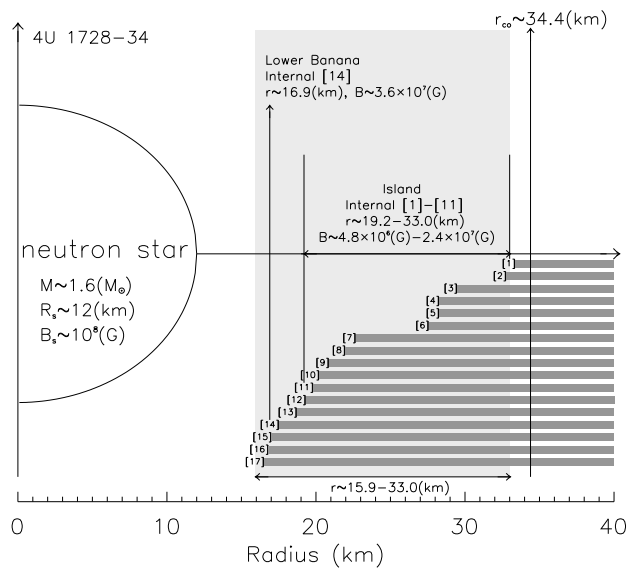}
\caption{Schematic diagram of the magnetosphere-disk structures for various positions in X-ray color-color diagram, where the number [1]-[17] are corresponding to the internals as described  in Figure \ref{r_b_mdot_rho}.
The position of the corotation radius and the ranges of the magnetic field strength are also shown.}
\label{schematic_diagram_r_r_co}
\end{figure}

Figure \ref{r_b_mdot_rho} (b) shows the plot of the $B(r)$ values at each interval in CCD, which
increases monotonically along the atoll track, from the island state to the banana state.
We also note that the $B(r)$ value at the island-banana gap shown in Figure \ref{r_b_mdot_rho}
(a) increases from $B(r)\sim8.6\times10^6$\,G in the
interval [6] to $B(r)\sim1.5\times10^7$\,G in the interval [7], doubling its value.
 While the $B(r)$ value at the vertex of the atoll track around
internal [15] in Figure \ref{r_b_mdot_rho} (a) is $B(r)\sim3.9\times10^7$\,G.
For clarity, the schematic diagram of the magnetosphere-disk structure in Figure \ref{schematic_diagram_r_r_co} also shows the range of the $B(r)$ values
of the interval [1]-[11] and interval [14], respectively.

\subsection{Corotation radius}

The corotation radius---$r_{\rm co}$ in NS-LMXB is the radial distance referring to the NS center where the accretion plasma corotates with the NS magnetosphere, meaning that the Keplerian orbital frequency $\nu_{\rm K}$ there equals the NS spin frequency $\nu_{\rm s}$ \citep{Bhattacharya91}.
Setting $\nu_{\rm K}=\nu_{\rm s}$ in equation (\ref{nu_K}), $r_{\rm co}$ can be derived as
\begin{equation}
\begin{aligned}
r_{\rm co}&=(\frac{GM}{4{\rm \pi}^2})^{1/3}\nu_{\rm s}^{-2/3}\\
&\approx34.4({\rm km})(\frac{M}{1.6\,{\rm M_\odot}})^{1/3}(\frac{\nu_{\rm s}}{363\,{\rm Hz}})^{-2/3}.
\end{aligned}
\label{r_co}
\end{equation}

We infer the corotation radius of 4U 1728-34 by equation (\ref{r_co}) with the inferred NS spin frequency of $\nu_{\rm s}\sim363$\,Hz \citep{Strohmayer96} and the assumed NS mass of $M\sim1.6\,{\rm M_\odot}$ \citep{Zhang11,Ozel16}, and obtain $r_{\rm co}\sim34.4$\,km (see Table \ref{QPOs}). We note that all the magnetosphere-disk radii ($r\sim15.9-33.0$\,km, see Table \ref{QPOs}) of the source inferred by the kHz QPOs are smaller than its corotation radius (see also Figure \ref{schematic_diagram_r_r_co}).

\section{Discussions and conclusions}
We assumed that the upper kHz QPO is ascribed to the Keplerian orbital motion and the NS magnetosphere is defined by the dipole magnetic field, as well as the detected kHz QPOs  from \citet{Di Salvo01} and NS spin (363\,Hz) inferred by the type I X-ray burst oscillation \citep{Strohmayer96}. Based on these assumptions, we investigated the evolution of the magnetosphere-disk structure of 4U 1728-34 along the atoll track. We find that the magnetosphere-disk radius of the source decreases monotonically from the island state to the banana state in CCD of the source. Below we summarize the details of the discussions and conclusions.

As the source evolves from the island state to the banana state, its magnetosphere-disk radius decreases nearly by half (see Table \ref{QPOs} and Figure \ref{r_b_mdot_rho} (a)), while the accretion disk gradually enters the stronger magnetic field region with the strength increasing by a factor of 9 (see see Table \ref{QPOs} and Figure \ref{r_b_mdot_rho} (b)). We note that these results rely on the simplistic assumptions of the Keplerian motion and magnetic dipole field (see Equations (\ref{nu_K})-(\ref{r_co})). However, the reality of the problem is much more complex: due to the viscous of the accretion plasma and its interaction with the NS magnetic field, the orbital velocity of the plasma at a certain radius may be less than the Keplerian orbital velocity there, and Equation (\ref{nu_K}) should be revised as
    \begin{equation}
    \nu_2=\xi\nu_{\rm K}=\xi\sqrt{\frac{GM}{4{\rm \pi}^2r^3}},
    \label{nu_K_R}
    \end{equation}
    where $0<\xi\leq1$. The value of $\xi$ can be estimated as follows: the observed maximal kHz QPO frequency of 4U 1728-34 is $\sim1200$\,Hz \citep{Migliari03}, inferring the Keplerian orbital radius of $r\sim15.5$\,km with the assumed NS mass of $1.6\,{\rm M_\odot}$. By assuming the actual emission radius should be less than 15.5 km, for example, $r\sim15$\,km, we can infer the value of $\xi$ to be
    \begin{equation}
    \xi=95\%(\frac{\nu_2}{1200\,{\rm Hz}})(\frac{M}{1.6\,{\rm M_\odot}})^{-1/2}(\frac{r}{15\,{\rm km}})^{3/2}.
    \label{xi}
    \end{equation}
    If $\xi$ is a constant, the variation of the magnetosphere-disk radius from the internal [1] to [17] can be derived as $r_{[1]}/r_{[17]}\propto(\nu_{2[1]}/\nu_{2[17]})^{-2/3}\sim(387/1158)^{-2/3}\approx2.1$. Furthermore, in vacuum, the dipole field lines are poloidal with $B\propto r^{-3}$, which increases from the internal [1] to [17] by a factor of $B_{[17]}/B_{[1]}\propto(r_{[1]}/r_{[17]})^{3}\sim(2.1)^{3}\approx9.2$. In the presence of the accretion disk, the field lines are initially "frozen-in" to the orbiting plasma, which hence are sheared in the $\phi$ direction \citep{Ghosh78,Shapiro83}. This shearing generates a sizeable $B_\phi$ component, causing the magnetic field strength around the accretion disk may be stronger than the dipole field, as a general form,
    \begin{equation}
    B(r)=B_{\rm s}(\frac{R}{r})^\alpha,
    \label{B_disk}
    \end{equation}
    with $\alpha\geq3$. If we assume $\alpha\sim3.5$, the magnetic field strength from the internal [1] to [17] should increase by a factor of $B_{[17]}/B_{[1]}\propto(r_{[1]}/r_{[17]})^{3.5}\sim(2.1)^{3.5}\approx13.4$. However, these corrections do not obviously affect the conclusions that, as the evolution of the atoll track of 4U 1728-34, the accretion disk moves toward the NS surface and gradually enters the stronger magnetic field region.

It can be seen from Figure \ref{r_b_mdot_rho} (a) that the whole atoll track of 4U 1728-34 is formed when the magnetosphere-disk radii are smaller than the corotation radius, i.e., $r/r_{\rm co}\propto(\nu_{\rm s}/\nu_2)^{2/3}\leq(363\,{\rm Hz}/387\,{\rm Hz})^{2/3}\sim0.95$. The accretion plasma corotates with the magnetosphere at $r_{\rm co}$, while according to the expression of the Keplerian orbital velocity:
    \begin{equation}
    \begin{aligned}
    v&=(2{\rm \pi}GM\nu_2)^{1/3}\\
    &\propto\nu_2^{1/3},
    \end{aligned}
    \label{Delt_v}
    \end{equation}
    as the accretion disk moves toward the NS surface from the internal [1] to [17], the orbital velocity of the plasma increases gradually by a factor of $(1158^{1/3}-387^{1/3})/387^{1/3}\sim44\%$. The large velocity difference $\Delta v\propto\nu_2^{1/3}-\nu_{\rm s}^{1/3}$ between the magnetosphere and the accretion disk may induce the strong interaction and release huge energy \citep{Wang17}. We suspect that this process can provide the particular physical environment to soften the photons, which causes the evolution of the atoll track from island state to banana state.

Figure \ref{schematic_diagram_r_r_co} shows a position gap around $r\sim27.2-22.3$\,km between the internal [6] and [7] (see Table \ref{QPOs}), which can be used to classify the magnetosphere-disk structures into two categories: (I)---internal [1]-[6] with $r\sim27.2-33.0$\,km, where the source is in the island state  and $r$ is near the corotation radius ($r_{\rm co}\sim34.4$\,km, see Table \ref{QPOs}); (II)---internal [7]-[17] with $r\sim15.9-22.3$\,km, where the source is in the island state and banana state with $r$ approaching the NS surface.
    This "gap" may be due to the lack of the observational time \citep{Zhang16}. However,
    we also suggest that the corotation radius and NS surface may be the two characteristic boundaries for the atoll track evolution: the corotation radius may be a soft boundary, and we guess the common rotation frequency of accretion plasma and magnetosphere may induce some type of resonance effect and release huge energy. Furthermore, the radiation around this position may be hard-photon dominated, as shown in the island state (see Figure \ref{r_b_mdot_rho}). While the NS surface is a hard boundary, and when the accretion plasma collide on this hard surface or on the local strong magnetic field \citep{Zhang06}, the source may release huge energy and produce much soft photons.
    This guess is supported by the fact that there is
    a vertex around the internal [15] in CCD, where the radiation varies from the hard photons dominated gradually to the soft photons dominated.
    We note from Figure \ref{schematic_diagram_r_r_co} that the inner radius corresponding to the vertex is quite near the NS surface ($r\sim16.4$\,km), and the distance difference of the inner radii between internal [15] ($r\sim16.4$\,km) and [17] ($r\sim15.9$\,km) is as short as $\sim0.5$\,km.

The parallel tracks in the kHz QPO evolution have been observed in 4U 1728-34 \citep{Mendez01,van der Klis01}, in other words, the same kHz QPO frequency could correspond to at least two different luminosities (or count rates) and colors during different observations. This phenomenon may deviate from our conclusion, since taking a different track may end up with different results for the same frequency. Here we have adopted the double accretion rates suggested by \citet{van der Klis01} to explain this phenomenon of parallel lines. As such, the luminosity of the source depends on the total mass accretion rate $\dot{M}_{\rm tot}$, which comes from the disk accretion $\dot{M}_{\rm d}$ and the radial accretion $\dot{M}_{\rm rad}$ (see also \citealt{Pan16}):
    \begin{equation}
    \dot{M}_{\rm tot}=\dot{M}_{\rm d}+\dot{M}_{\rm rad}.
    \label{Mdot}
    \end{equation}
    The QPOs are likely to origin from the accretion disk \citep{van der Klis01} due to the quasi-periodic property and are affected by $\dot{M}_{\rm d}$. As  $\dot{M}_{\rm d}$ increases, the accretion disk moves toward the NS surface, accompanied by the increasing of the kHz QPO frequency and the formation of the atoll track. In addition, the radial accretion may exist, since the accretion disk is thickened by the radiation pressure as it moves toward the NS surface, which causes an approximate spherical accretion in the region that is smaller than the corotation radius of $r_{\rm co}\sim34.4$\,km. The different radial accretion and
    the disk accretion rates can account for the same magnetosphere-disk radii, which can account for the same kHz QPOs, while they produce the different luminosity. In other words,  $\dot{M}_{\rm rad}$ can modulate both the luminosity and the colors of the source, which modulates a same magnetosphere-disk radius as a  $\dot{M}_{\rm d}$ does, thus a phenomenon  of the parallel tracks occurs.

Based on the above discussions, we suggest that the atoll track in CCD of 4U 1728-34  may correspond to
the particular  magnetosphere-disk boundary, where the corotation radius and NS hard surface could
play a significant role in the different radiation processes. In addition,
 it seems that the magnetic field strength and spin-up state of NS may also contribute to
  the radiation composition in the CCD.  As a further exploration, however, similar investigations in other atoll sources and Z sources are needed to test whether these phenomena
   are common to all NS-LMXBs. If confirmed, these results will help to understand
    the formation mechanism of the atoll CCD track, to probe the accretion environment and the magnetic field structure
     around NS, and also to test the presence mechanism  of kHz QPOs.

\begin{acknowledgements}
This work is supported by the National Program on Key Research and Development Project (Grant No. 2016YFA0400803),
the National Natural Science Foundation of China (Grant No. 11703003, No. 11673023 and No. U1731238).
\end{acknowledgements}

\begin{thebibliography}{}

\bibitem[\protect\citeauthoryear{Abramowicz et al.}{2003}]{Abramowicz03b}
Abramowicz, M. A., Karas, V., Klu\'zniak, W., Lee, W. H., \& Rebusco, P. 2003, PASJ, 55, 467
%
\bibitem[\protect\citeauthoryear{Altamirano et al.}{2008}]{Altamirano08b}
Altamirano, D., van der Klis, M., Me\'ndez, M., et al. 2008, ApJ, 685, 436
%
\bibitem[\protect\citeauthoryear{Belloni et al.}{2002}]{Belloni02}
Belloni, T., Psaltis, D., \& van der Klis, M. 2002, ApJ, 572, 392
%
\bibitem[\protect\citeauthoryear{Belloni et al.}{2005}]{Belloni05}
Belloni, T., M\'endez, M., \& Homan, J. 2005, A\&A, 437, 209

\bibitem[\protect\citeauthoryear{Belloni et al.}{2007}]{Belloni07}
Belloni, T., M\'endez, M., \& Homan, J. 2007, MNRAS, 376, 1133
%
\bibitem[\protect\citeauthoryear{Bhattacharya \& van den Heuvel}{1991}]{Bhattacharya91}
Bhattacharya, D., \& van den Heuvel, E. P. J. 1991, Physics Reports, 203, 1
%
\bibitem[\protect\citeauthoryear{Boutloukos \& Lamb}{2008}]{Boutloukos08a}
Boutloukos, S., \& Lamb, F. K. 2008, in Bassa, C. G. et al., eds, 40 Years of Pulsars:
Millisecond Pulsars, Magnetars, and More, AIP Conf. Ser., Vol. 983., Am. Inst. Phys., Melville, NY, p. 533
%
\bibitem[\protect\citeauthoryear{Burderi \& Di Salvo}{2013}]{Burderi13}
Burderi, L., \& Di Salvo, T. 2013, Memorie della Societa Astronomica Italiana, 84, 117
%
\bibitem[\protect\citeauthoryear{Di Salvo et al.}{2001}]{Di Salvo01}
Di Salvo, T., M\'endez, M., van der Klis, M., Ford, E., \& Robba, N. R. 2001, ApJ, 546, 1107
%
\bibitem[\protect\citeauthoryear{Ford \& van der Klis}{1998}]{Ford98}
Ford, E. C., \& van der Klis, M. 1998, ApJ, 506, L39
%
\bibitem[\protect\citeauthoryear{Ford et al.}{2000}]{Ford00}
Ford, E. C., van der Klis, M., M\'endez, M., et al. 2000, ApJ, 537, 368
%
\bibitem[\protect\citeauthoryear{Fridriksson et al.}{2015}]{Fridriksson15}
Fridriksson, J. K., Homan, J., \& Remillard, R. A. 2015, ApJ, 809, 52
%
\bibitem[\protect\citeauthoryear{Ghosh \& Lamb}{1978}]{Ghosh78}
Ghosh, P. \& Lamb, F. K., 1978, ApJ, 223, L83

%
\bibitem[\protect\citeauthoryear{Harding}{2013}]{Harding13}
Harding, A. K. 2013, Frontiers of Physics, 8, 679
%
\bibitem[\protect\citeauthoryear{Hasinger \& van der Klis}{1989}]{Hasinger89}
Hasinger, G., \& van der Klis, M. 1989, A\&A, 225, 79
%
\bibitem[\protect\citeauthoryear{Hasinger et al.}{1990}]{Hasinger90a}
Hasinger, G., van der Klis, M., Ebisawa, K., Dotani, T., \& Mitsuda, K. 1990, A\&A, 235, 131
%
\bibitem[\protect\citeauthoryear{Hasinger}{1990}]{Hasinger90b}
Hasinger, G. 1990, Reviews in Modern Astronomy, 3, 60
%
\bibitem[\protect\citeauthoryear{Homan et al.}{2002}]{Homan02}
Homan, J., van der Klis, M., Jonker, P. G., et al. 2002, ApJ, 568, 878
%
\bibitem[\protect\citeauthoryear{Homan et al.}{2007}]{Homan07}
Homan, J., van der Klis, M., Wijnands, R., et al. 2007, ApJ, 656, 420
%
\bibitem[\protect\citeauthoryear{Homan et al.}{2010}]{Homan10}
Homan, J., van der Klis, M., Fridriksson, J. K., et al. 2010, ApJ, 719, 201
%
\bibitem[\protect\citeauthoryear{Jonker et al.}{2002}]{Jonker02}
Jonker, P. G., van der Klis, K., Homan, J., et al. 2002, MNRAS, 333, 665
%
\bibitem[\protect\citeauthoryear{Kaaret et al.}{1998}]{Kaaret98}
Kaaret, P., Yu, W., Ford, E. C., \& Zhang, S. N. 1998, ApJ, 497, L93
%
\bibitem[\protect\citeauthoryear{Klu\'zniak et al.}{1990}]{Kluzniak90}
Klu\'zniak, W., Michelson, P., \& Wagoner, R. V. 1990, ApJ, 358, 538
%
\bibitem[\protect\citeauthoryear{Lamb \& Miller}{2001}]{Lamb01}
Lamb, F. K., \& Miller, M. C. 2001, ApJ, 554, 1210
%
\bibitem[\protect\citeauthoryear{Lin et al.}{2009}]{Lin09}
Lin, D. C., Remillard, R. A., \& Homan, J. 2009, ApJ, 696, 1257
%
\bibitem[\protect\citeauthoryear{Lin et al.}{2012}]{Lin12}
Lin, D. C., Remillard, R. A., Homan, J., \& Barret, D. 2012, ApJ, 756, 34
%
\bibitem[\protect\citeauthoryear{M\'endez \& van der Klis}{1999}]{Mendez99a}
M\'endez, M., \& van der Klis, M. 1999, ApJ, 517, L51
%
\bibitem[\protect\citeauthoryear{M\'endez et al.}{1999}]{Mendez99}
M\'endez, M., van der Klis, K., Ford, E. C., Wijnands, R., \& van Paradijs, J. 1999, ApJ, 511, L49
%
\bibitem[\protect\citeauthoryear{M\'endez et al.}{2001}]{Mendez01}
M\'endez, M., van der Klis, K., \& Ford, E. C. 2001, ApJ, 561, 1016
%
\bibitem[\protect\citeauthoryear{M\'endez et al.}{2006}]{Mendez06}
M\'endez, M. 2006, MNRAS, 371, 1925
%
\bibitem[\protect\citeauthoryear{Migliari et al.}{2003}]{Migliari03}
Migliari, S., van der Klis, M., Fender, R. P. 2003, MNRAS, 345, L35
%
\bibitem[\protect\citeauthoryear{Miller et al.}{1998}]{Miller98}
Miller, M. C., Lamb, F. K., \& Psaltis, D. 1998, ApJ,  508, 791
%
\bibitem[\protect\citeauthoryear{Miller \& Miller}{2015}]{Miller15}
Miller, M. C., \& Miller, J. M. 2015, Physics Reports, 548, 1
%
\bibitem[\protect\citeauthoryear{Mondal et al.}{2017}]{Mondal17}
Mondal, A. S., Pahari, M., Dewangan, G. C., Misra, R., \& Raychaudhuri, B. 2017, MNRAS, 466, 4991
%
\bibitem[\protect\citeauthoryear{Mukherjee}{2012}]{Mukherjee12}
Mukherjee, A., \& Bhattacharyya, S. 2012, ApJ, 756, 55
%
\bibitem[\protect\citeauthoryear{\"Ozel \& Freire}{2016}]{Ozel16}
\"Ozel, F., \& Freire, P. 2016, ARA\&A, 54, 401
%
\bibitem[\protect\citeauthoryear{Pan et al.}{2016}]{Pan16}
Pan, Y. Y., Song, L. M., Zhang, C. M., \& Tong, H. 2016, MNRAS, 461, 2
%
\bibitem[\protect\citeauthoryear{Psaltis et al.}{1999}]{Psaltis99}
Psaltis, D., Belloni, T., \& van der Klis, M. 1999, ApJ, 520, 262 
%
\bibitem[\protect\citeauthoryear{Shaposhnikov et al.}{2003}]{Shaposhnikov03}
Shaposhnikov, N., Titarchuk, L., \& Haberl, F. 2003, ApJ, 593, 35
%
\bibitem[\protect\citeauthoryear{Shapiro \& Teukolsky}{1983}]{Shapiro83}
Shapiro, S. L., \& Teukolsky S. A. 1983, Black Holes, White Dwarfs, and Neutron Stars: The Physics of Compact Objects. Wiley Interscience, New York
%
\bibitem[\protect\citeauthoryear{Stella \& Vietri}{1999}]{Stella99a}
Stella, L., \& Vietri, M. 1999, Phys. Rev. Lett., 82, 17 
%
\bibitem[\protect\citeauthoryear{Stella et al.}{1999}]{Stella99b}
Stella, L., Vietri, M., \& Morsink, S. M. 1999, ApJ, 524, L63
%
\bibitem[\protect\citeauthoryear{Strohmayer et al.}{1996}]{Strohmayer96}
Strohmayer, T. E., Zhang, W., Swank, J. H., et al. 1996, ApJ, 469, L9
%
\bibitem[\protect\citeauthoryear{Tarana et al.}{2011}]{Tarana11}
Tarana, A., Belloni, T., Bazzano, A., M\'endez, M., \& Ubertini, P. 2011, MNRAS, 416, 873
%
\bibitem[\protect\citeauthoryear{van der Klis}{2000}]{van der Klis00}
van der Klis, M. 2000, ARA\&A, 38, 717
%
\bibitem[\protect\citeauthoryear{van der Klis}{2001}]{van der Klis01}
van der Klis, M. 2001, ApJ, 561, 943
%
\bibitem[\protect\citeauthoryear{van der Klis}{2006}]{van der Klis06}
van der Klis, M. 2006, in Lewin, W. H. G., van der Klis, M., eds,
Compact Stellar X-Ray Sources, Cambridge Univ. Press, Cambridge, p.39
%
\bibitem[\protect\citeauthoryear{van der Klis}{2016}]{van der Klis16}
van der Klis, M. (EWASS) 2016, invited talk in European Week of Astronomy and Space Science,
Timing Low-Mass X-Ray Binaries and Accreting Millisecond Pulsars, 4-8 July 2016, Athens, Greece
%
\bibitem[\protect\citeauthoryear{van Straaten et al.}{2000}]{van Straaten00}
van Straaten, S., Ford, E. C., van der Klis, M., M\'endez, M., \& Kaaret, P. 2000, ApJ, 540, 1049 
%
\bibitem[\protect\citeauthoryear{van Straaten et al.}{2002}]{van Straaten02}
van Straaten, S., van der Klis, M., Di Salvo, T., \& Belloni, T. 2002, ApJ, 568, 912 
%
\bibitem[\protect\citeauthoryear{van Straaten et al.}{2003}]{van Straaten03}
van Straaten, S., van der Klis, M., \& M\'endez, M. 2003, ApJ, 596, 1155 
%
\bibitem[\protect\citeauthoryear{Wang et al.}{2014}]{Wang14}
Wang, D. H., Chen, L., Zhang, C. M., \& Qu, J. L. 2014, Astron. Nachr., 335, 168
%
\bibitem[\protect\citeauthoryear{Wang et al.}{2015}]{Wang15}
Wang, D. H., Chen, L., Zhang, C. M., et al. 2015, MNRAS, 454, 1231
%
\bibitem[\protect\citeauthoryear{Wang et al.}{2017}]{Wang17}
Wang, D. H., Zhang, C. M., Lei, Y. J., et al. 2017, MNRAS, 1111, 1117
%
%
\bibitem[\protect\citeauthoryear{Wang et al.}{2018}]{Wang18}
Wang, D. H., Zhang, C. M., Qu, J. L., \& Yang, Y. Y. 2018, MNRAS, 473, 4862
%
\bibitem[\protect\citeauthoryear{Wijnands et al.}{1997}]{Wijnands97}
Wijnands, R., Homan, J., van der Klis, M., et al. 1997, ApJ, 490, L157
%
\bibitem[\protect\citeauthoryear{Wijnands et al.}{1998}]{Wijnands98}
Wijnands, R., Me\'ndez, M., van der Klis, M., et al. 1998, ApJ, 504, L35
%
\bibitem[\protect\citeauthoryear{Zhang}{2004}]{Zhang04}
Zhang, C. M. 2004, A\&A, 423, 401
%
\bibitem[\protect\citeauthoryear{Zhang et al.}{2006}]{Zhang06a}
Zhang, C. M., Yin, H. X., Zhao, Y. H., Zhang, F., \& Song, L. M. 2006, MNRAS, 366, 1373
%
\bibitem[\protect\citeauthoryear{Zhang \& Kojima}{2006}]{Zhang06}
Zhang, C. M., \& Kojima, Y. 2006, MNRAS, 366, 137
%
\bibitem[\protect\citeauthoryear{Zhang et al.}{2011}]{Zhang11}
Zhang, C. M., Wang, J., Zhao, Y. H., et al. 2011, A\&A, 527, 83
%
\bibitem[\protect\citeauthoryear{Zhang \& Wang}{2013}]{Zhang13}
Zhang, C. M., \& Wang, D. H. 2013, in Zhang, C. M., Belloni, T., M\'endez, M. et al., eds,
Feeding Compact Objects: Accretion on All Scales, Proceedings of IAU Symp. 290,
Cambridge: Cambridge University Press, pp. 381-385
%
\bibitem[\protect\citeauthoryear{Zhang et al.}{2016}]{Zhang16}
Zhang, G., M\'endez, M., Zamfir, M., \& Cumming, A. 2016, MNRAS, 455, 2004


\end{thebibliography}
\end{document}